\documentclass[aps,prc,twocolumn,showpacs,preprintnumbers,nofootinbib,float,superscriptaddress,longbibliography]{revtex4-1}
\usepackage{graphicx}
\usepackage{color}
\usepackage[dvipsnames]{xcolor}
\usepackage[colorlinks=true, pdfstartview=FitV, linkcolor=RedOrange, citecolor=ForestGreen, urlcolor=blue]{hyperref}
\usepackage{amsmath}
\usepackage{amssymb}
\usepackage{bm}
\usepackage{physics}
\usepackage{placeins}

\begin{document}

\title{A Gaussian Process Generative Model for QCD Equation of State}

\author{Jiaxuan Gong} \email{jacksongong2017@gmail.com}
\affiliation{Northville High School, Northville, Michigan 48618, USA}
\affiliation{Department of Physics and Astronomy, Wayne State University, Detroit, Michigan 48201, USA}
\author{Hendrik Roch} \email{Hendrik.Roch@wayne.edu}
\affiliation{Department of Physics and Astronomy, Wayne State University, Detroit, Michigan 48201, USA}
\author{Chun Shen} \email{chunshen@wayne.edu}
\affiliation{Department of Physics and Astronomy, Wayne State University, Detroit, Michigan 48201, USA}

\begin{abstract}
We develop a generative model for the nuclear matter equation of state at zero net baryon density using the Gaussian Process Regression method. We impose first-principles theoretical constraints from lattice QCD and hadron resonance gas at high- and low-temperature regions, respectively. By allowing the trained Gaussian Process Regression model to vary freely near the phase transition region, we generate random smooth cross-over equations of state with different speeds of sound that do not rely on specific parameterizations.
We explore a collection of experimental observable dependencies on the generated equations of state, which paves the groundwork for future Bayesian inference studies to use experimental measurements from relativistic heavy-ion collisions to constrain the nuclear matter equation of state.
\end{abstract}

\maketitle

\section{Introduction}
\label{sec:intro}
One of the primary goals of relativistic heavy-ion collisions is the determination of the equation of state (EOS) of the hot and dense matter of Quantum Chromodynamics (QCD) produced in these collisions~\cite{Pratt:2015zsa, Busza:2018rrf, Gardim:2019xjs, Shen:2020mgh, An:2021wof, Monnai:2021kgu, Almaalol:2022xwv, Sorensen:2023zkk, Achenbach:2023pba, Arslandok:2023utm, Du:2024wjm}. The equation of state maps out the detailed structure of the nuclear matter phase diagram, probed experimentally by varying the center-of-mass energy of the collisions.
At vanishing baryon chemical potential $\mu_B$, the EOS can be computed from first principles using Lattice QCD techniques~\cite{Borsanyi:2013bia, HotQCD:2014kol}. These results have become standard inputs for modeling the dynamics of high-energy heavy-ion collisions at the Large Hadron Collider (LHC)~\cite{Shen:2014vra, Giacalone:2017dud, Putschke:2019yrg, Schenke:2020mbo, Nijs:2020ors, Hirvonen:2024zne}.
At finite net baryon density, despite many advancements in new techniques and computing~\cite{Fodor:2004nz, Gavai:2008zr, Bazavov:2017dus, Giordano:2020roi, Borsanyi:2021sxv}, the lattice QCD calculation of the EOS remains a very challenging task because of the sign problem~\cite{Ratti:2018ksb}. 
On the other hand, precision measurements from the Beam Energy Scan (BES) program at the Relativistic Heavy Ion Collider (RHIC) provide opportunities to explore different temperatures and non-vanishing densities~\cite{Caines:2009yu,Mohanty:2011nm,Mitchell:2012mx,Odyniec:2015iaa}.
The experimental measurements at different energies can shed light on the transition between hadronic matter and Quark-Gluon Plasma (QGP), allowing searching for a critical point and the associated first-order phase transition line in the QCD phase diagram at finite net baryon density (see Refs.~\cite{STAR:2010vob,Luo:2017faz,Bzdak:2019pkr,Shen:2020mgh,An:2021wof}).

Most of the phenomenological studies~\cite{Auvinen:2017fjw, Sun:2022rjh, Shen:2022oyg, Shen:2023aeg, Shen:2023awv, Jiang:2023fai, Du:2022yok, Du:2023gnv, Du:2023msv, Shen:2023pgb, Plumberg:2024leb, Pihan:2024lxw, Monnai:2024pvy, Jahan:2024wpj} at the RHIC BES program employed a lattice QCD EOS based on Taylor expansion schemes to extrapolate to finite $\mu_B$ regions~\cite{Monnai:2019hkn, Noronha-Hostler:2019ayj, Fotakis:2019nbq, Karthein:2021nxe, Mondal:2021jxk}.
With the recent developments in Bayesian inference studies in relativistic heavy-ion collisions~\cite{Novak:2013bqa, Bernhard:2019bmu, Nijs:2020ors, JETSCAPE:2020mzn}, determination of the QCD equation of state from experimental measurements was pioneered in Ref.~\cite{Pratt:2015zsa}. Because of the thermodynamics constraints, varying the speed of sound in QCD equations of state is usually complex and not easy to generalize to finite densities~\cite{Pratt:2015zsa, Giacalone:2023cet}.

In this work, we introduce a machine-learning-based method to generate physical random sets of equations of state using the Gaussian Process Regression (GPR). 
To explore this method, we will provide a first case study at zero net baryon density, where the thermodynamic pressure $P$ is only a function of temperature ($T$).
In this case, the real QCD EOS is known from first-principle Lattice QCD calculations, which allows us to study the effect of the randomly generated EOS sets on final-state observables in heavy-ion collisions and compare them to those obtained from the lattice QCD EOS.
Such a study at zero net baryon density can provide useful insights into the temperature range that bulk observables can probe about the QCD equation of state.
Our approach can be easily extended to finite densities,~e.g., $P(T, \mu_B)$, and incorporate various theoretical constraints at different phase diagram regions.

In Sec.~\ref{sec:EOS_generator}, we will introduce the setup for the Gaussian Process Regression with theoretical constraints. Using the trained GPR as a generative model, we apply filters to check thermodynamic relations and generate physical equations of state. In Sec.~\ref{sec:results}, we select two equations of state from the generated set and incorporate them in the \texttt{iEBE-MUSIC} framework for large-scale simulations for heavy-ion collisions. We compare the final-state flow observables with those from the lattice QCD EOS. Section~\ref{sec:conclusion_outlook} concludes our study.

\section{The Equation of State Generator}
\label{sec:EOS_generator}

In this section, we will discuss the setup of the Gaussian Process Regression model, train it with theoretical constraints from first principles, and generate random sets of EOS. We chose the Gaussian Process Regression as a generative model because it offers a non-parametric framework for generating random functions with well-defined derivatives. 
Similar approaches were used in modeling the neutron star EOS~\cite{Landry:2018prl, Mroczek:2023zxo}.
In this work, we restrict our scope to the smooth cross-over type of equation of state at zero net baryon chemical potential. One can introduce special kernel functions in the Gaussian Process to mimic a first-order phase transition~\cite{Mroczek:2023zxo}. But we will leave it for future work.
A helpful general reference about the Gaussian Process (GP) can be found in Ref.~\cite{williams2006gaussian}.

\subsection{Gaussian Process Regression}
\label{subsec:GPR}

A Gaussian Process model describes the probability distribution over possible functions that fit a set of observation points. It provides a flexible and compact representation for making regression predictions for the function and its uncertainty.

For a given set of points of a function $f(\theta)$, namely $\{\theta_i, f_i\}$ ($f_i \equiv f(\theta_i)$), the Gaussian Process Regression models the probability as a multi-variant Gaussian distribution,
\begin{align}
    P(\{f_i\} | \{\theta_i\}) = \mathrm{MVN}(\{f_i\} | \{\mu_i\}, K)
\end{align}
with
\begin{align}
    &\mathrm{MVN}(\{f_i\} | \{\mu_i\}, K) = \nonumber \\
    & \qquad \frac{1}{(2\pi)^{D/2} |K|^{1/2}} \exp\left[-\frac{1}{2} (f_i - \mu_i)^\mathsf{T} K^{-1} (f_i - \mu_i) \right].
\end{align}
Here, $\mu_i(\theta_i)$ is the mean function and the variance matrix $K$ is defined by a kernel function for its element $K_{ij} = k(\theta_i, \theta_j)$. The dimension $D$ equals the number of observation points in the $\{\theta_i, f_i\}$ set. Under the assumption of a zero-mean distribution, $\mu_i = 0$, the GPR makes the prediction for a function $f$ at a point $\theta^*$,
\begin{align}
    f(\theta^*) &= K(\theta^*, \theta_k) K^{-1}(\theta_k, \theta_j) f_j,
\end{align}
where repeated indices $j$ and $k$ sum over all the observation points. The covariance matrix between two prediction points at $\theta^*_i$ and $\theta^*_j$ is
\begin{align}
    K_{ij} &= k(\theta^*_i, \theta^*_j) - k(\theta^*_i, \theta_m) K^{-1}(\theta_m, \theta_n) k(\theta_n, \theta^*_j),
\end{align}
where the indices $m$ and $n$ sum over all the observables points.

In this work, we use a standard squared-exponential (aka radial basis function (RBF)) kernel,
\begin{align}
    k(\theta_i, \theta_j) = \sigma^2 \exp\left(-\frac{d(\theta_i, \theta_j)^2}{2l^2}\right)
    \label{eq:GPkernel}
\end{align}
with the Euclidean distance measure $d(\theta_i,\theta_j) = \vert \theta_i - \theta_j\vert$. The hyperparameters $\sigma$ and $l$ control the functions' behavior: $\sigma$ sets the overall strength of the correlations, and $l$ governs the length scale over which they occur. We employ the Gaussian Process implemented in the \texttt{scikit-learn} Python package.

\subsection{Non-parametric representation of the equation of state}

At zero net baryon density, the QCD EOS is a one-dimensional function $P(T)$ that relates the thermal pressure with local temperature. We define the scaled pressure variable $\tilde{P} \equiv P/T^4$ as a unitless quantity. To ensure the pressure stays positive from the generative model, we build a GP for $\ln(\tilde{P})$ as a function of $\ln(T/T_0)$ with $T_0 = 1$~GeV. We will impose thermodynamic and causality constraints as filters for the generated EOS sets.

To train the GPR, we include low-temperature data ($T\leq 0.13\;\mathrm{GeV}$) from the Hadron Resonance Gas (HRG) model and high-temperature data ($T\geq 0.7\;\mathrm{GeV}$) from and Lattice QCD calculations~\cite{HotQCD:2014kol}. We set $\sigma = 1$ for the kernel in Eq.~\eqref{eq:GPkernel}, and the optimized length scale is $l = 0.487$\footnote{One can allow the hyper-parameter $\sigma$ to vary in the optimization. We did not find a noticeable improvement in the overall fit quality. Therefore, we leave this parameter to be fixed to unity, which is a popular choice for standard Gaussian Process Regression.}. This $l$ value translates to a correlation length of $\sim$120 MeV at $T = 200$\,MeV. It means that our GPR does not overfit the training data and allows for substantial variations of the reduced pressure over a temperature range of 100 MeV.

Once the GPR is trained, we can sample random instances from the GPR as different sets of EOS, $P(T)$. Then, we compute the entropy and speed of sound for the EOS and reject those that violate the thermodynamic and causality constraints. 

\subsection{Thermodynamic relations and physics constraints}
\label{subsec:thermodynamics}

Provided the EOS as $P(T)$, we can compute other thermodynamic properties like energy density $e$ and entropy density $s$ by differentiating $P(T)$ as follows,
\begin{align}
    s &= \frac{\partial P}{\partial T},\\
    e &= Ts - P,\\
    c_s^2 &= \frac{\partial P}{\partial e} = \frac{\partial P/\partial T}{\partial e/\partial T} = \frac{\partial P/\partial T}{T \partial^2 P/\partial^2 T}\label{eq:cs2}.
\end{align}

All randomly generated EOS curves have to fulfill at each value of $T$ the following inequalities imposed by thermodynamics and causality:
\begin{align}
    &\frac{\partial P}{\partial T}>0, \\
    &\frac{\partial^2 P}{\partial T^2}>0\;\text{(compressibility)},\\
    &c_s^2 \ge 0 \;\text{and}\; c_s^2 < 1\;\text{(causality)}.
\end{align}
A recent work~\cite{Hippert:2024hum} proposed that $c_s^2 < 0.781$ in relativistic transient hydrodynamics.
We apply an upper limit on the speed of sound $c_s^2 < c_{s,\mathrm{max}}^2$ before deploying them to simulations for relativistic heavy-ion collisions.
%at zero net baryon density since most of the theories give $c_s^2 \le 1/3$ without chemical potential~\cite{Moore:2024gmt}.
In this work, we choose $c_{s,\mathrm{max}}^2 = 1/2$ for a demonstration. We chose it to be larger than the ideal gas limit ($c_s^2 = 1/3$) to test whether our numerical simulations can handle EOS with a local speed of sound larger than the ideal gas limit. See more discussion related to causality in Sec.~\ref{subsec:transport_coeff}.

\subsection{EOS generator in action}
\label{subsec:generated_EOS_curves}

With all the ingredients, we can now generate random EOS curves from our GPR.

\begin{figure*}[t!]
    \centering
    \includegraphics[width=0.49\textwidth]{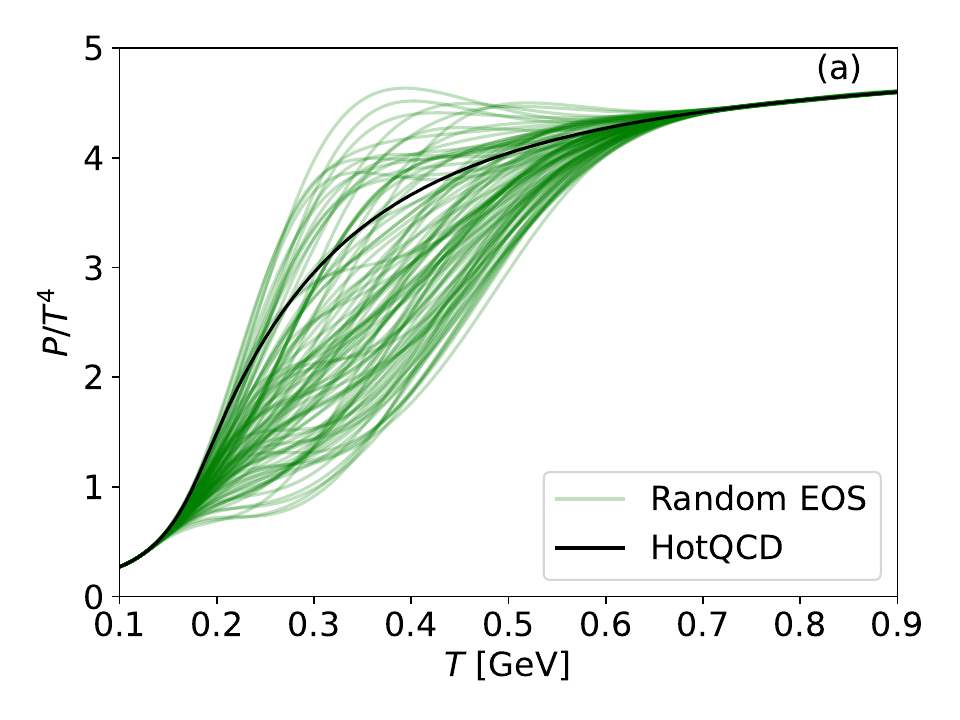}
    \includegraphics[width=0.49\textwidth]{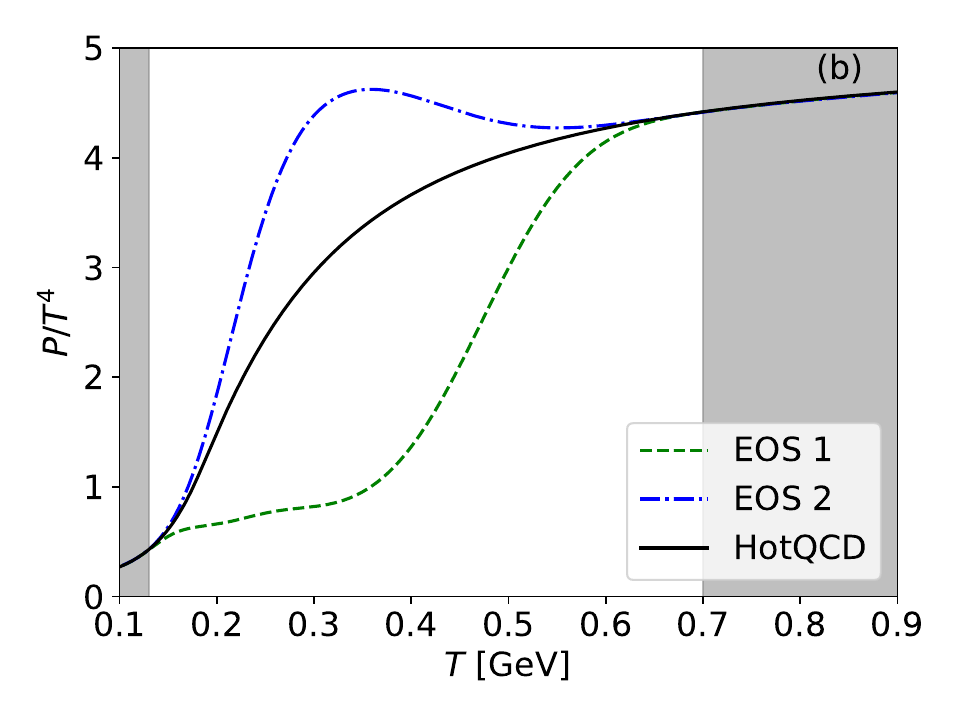}
    \includegraphics[width=0.49\textwidth]{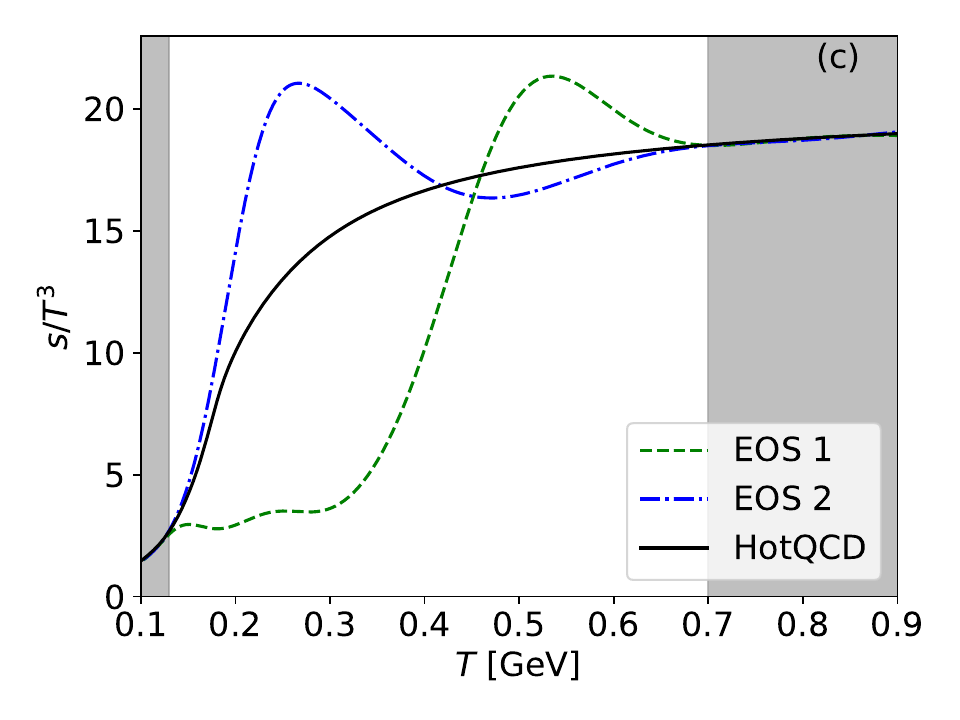}
    \includegraphics[width=0.49\textwidth]{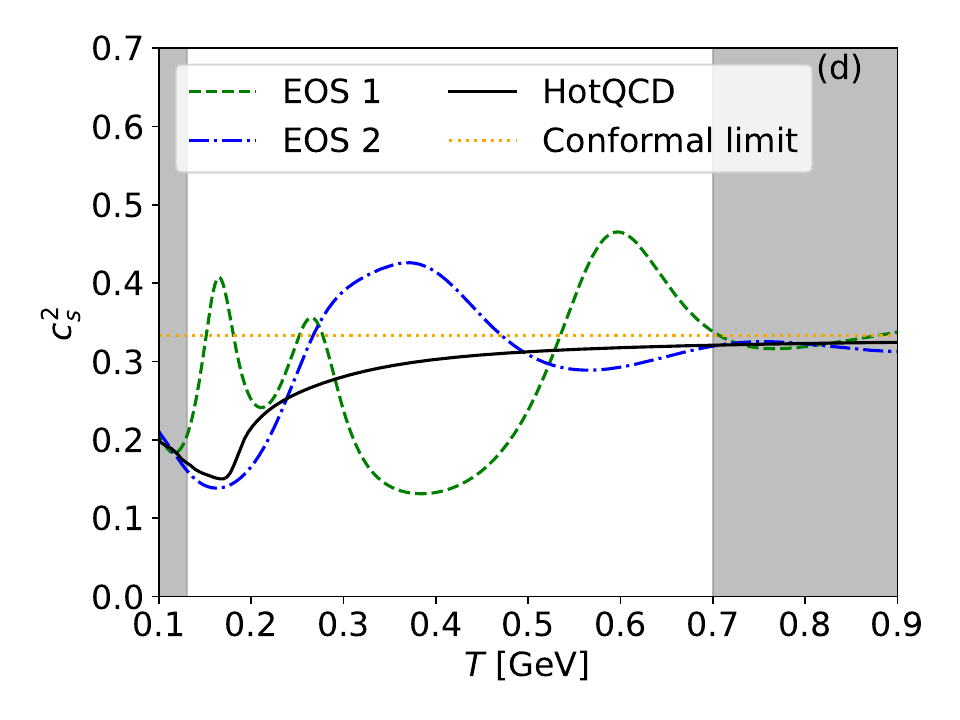}
    \caption{The scaled pressure $P/T^4$ for randomly generated EOSs from our GPR model (a). The other panels show $P/T^4$ (b), $s/T^3$ (c), and the square of the speed of sound $c_s^2$ (d) vs. temperature $T$ for two sets of EOS. The HotQCD + HRG EOS~\cite{HotQCD:2014kol, Moreland:2015dvc} (the black solid line) is shown as a reference. The gray regions are used to fit the GPR.}
    \label{fig:EOS}
\end{figure*}

Figure~\ref{fig:EOS}a shows 100 randomly generated equations of state for their scaled pressure compared to the HotQCD + HRG EOS parameterized in Ref.~\cite{Moreland:2015dvc} (the full black line). The GPR is allowed to vary freely at all temperature values in between the constraints. The randomly generated EOS sets show a wide spread around the lattice EOS in the unconstrained region and become more and more constrained towards the low and high-temperature region, as expected. 

To explore the maximal effect of the EOS on the final state observables in relativistic heavy-ion collisions, we select two sets of EOS that envelope all the randomly generated EOS sets. To make the selection more efficient, we introduced anchor points at $(T, \tilde{P}) = (0.25\;\mathrm{GeV}, 0.75)$ for EOS 1 and at $(0.25\;\mathrm{GeV}, 3.5)$ for EOS 2. Their scaled pressures as functions of temperature are shown in Fig.~\ref{fig:EOS}b. Although EOS 2 shows non-monotonic dependence for its scaled pressure on temperature, its thermal pressure is a monotonic function of temperature. 

Figure~\ref{fig:EOS}c shows the entropy density normalized by $T^3$ as a function of temperature. Compared to the lattice QCD EOS, the normalized entropy from the two sets of EOS at the extreme shows a non-monotonic dependence on temperature. Lastly, Figure~\ref{fig:EOS}d shows the square of the speed of sound $c_s^2$ for different EOS. Compared to the lattice QCD EOS, EOS 1 with smaller $P/T^4$ and $s/T^3$ near the phase cross-over region has a larger speed of sound in this temperature region. Because we impose the constraints on values of $P/T^4$ at high temperatures to those of the lattice QCD EOS, the speed of sound of the EOS oscillates around that of the lattice QCD EOS. The stiffer EOS near the phase cross-over region has a lower speed of sound at higher temperatures and vice versa.

\section{Deploy EOS to heavy-ion simulations}
\label{sec:results}

With the randomly generated equations of state, we invert $P(T)$ and $e(T)$ to tabulate thermal pressure as a function of energy density $P(e)$, which is then an input for hydrodynamic simulations to model the dynamical evolution of relativistic heavy-ion collisions. 

\subsection{Setup event-by-event simulations}
\label{subsec:EBEsimulations}
To study the effects of different EOS on final state observables, we employ the \texttt{iEBE-MUSIC} framework~\cite{Schenke:2020mbo, Shen:2022oyg} to simulate event-by-event relativistic heavy-ion collisions.

In this work, we perform (2+1)D boost-invariant simulations with the IP-Glasma + MUSIC + UrQMD model within the \texttt{iEBE-MUSIC} framework~\cite{Schenke:2020mbo}. We set the model parameters based on a recent work~\cite{Mantysaari:2024uwn}.

The initial states of the heavy-ion collisions are simulated event-by-event using the \texttt{IP-Glasma} model ~\cite{Schenke:2012wb, Mantysaari:2017cni}. The Glasma pre-equilibrium evolution is propagated to $\tau_{\rm hydro}=0.4\;\mathrm{fm}/c$, at which the system's energy-momentum tensor is mapped to hydrodynamic fields using the Landau matching condition, $T^{\mu\nu} u_\nu = e u^\mu$~\cite{Schenke:2019pmk, Schenke:2020mbo}. 
After this time, the energy-momentum tensor is evolved using second-order relativistic viscous hydrodynamics based on the Denicol-Niemi-Molnar-Rischke (DNMR) theory~\cite{Denicol:2012cn}. We incorporate the generated EOS from the GPR in solving the equation of motion of hydrodynamics in \texttt{MUSIC}~\cite{Schenke:2010nt, Schenke:2011bn, Paquet:2015lta}. When the medium local energy density drops to the switching energy density $e_\mathrm{sw} = 0.18$ GeV/fm$^3$, the fluid cells on the constant energy surface are identified using the \texttt{CORNELIUS} algorithm~\cite{Huovinen:2012is} and they are converted to particles according to the Cooper-Frye prescription~\cite{Cooper:1974mv}. Out-of-equilibrium ($\delta f$) corrections to particle emission are considered using the Grad's moment method~\cite{Shen:2014vra, Zhao:2022ugy}. The thermally emitted hadrons are fed to a hadronic transport model, \texttt{UrQMD}~\cite{Bass:1998ca,Bleicher:1999xi}, to model the non-equilibrium hadronic scatterings and resonance decays in the dilute hadronic phase.

\subsection{Second-order transport coefficients and causality constraints}
\label{subsec:transport_coeff}

The DNMR hydrodynamic theory introduces the second-order transport coefficients, such as the relaxation times for shear stress tensor and bulk viscous pressure. They are usually parameterized as functions of the specific shear and bulk viscosities~\cite{Denicol:2014vaa}. The parameterization for bulk viscous relaxation time $\tau_\Pi$ is chosen to depend on $(1/3 - c_s^2)$~\cite{Kanitscheider:2009as, Rougemont:2017tlu, Denicol:2014vaa}, where $c_s^2$ is the speed of sound squared from the EOS. Since we allow $c_s^2 \geq 1/3$ in our GPR generator, we need to modify the conventional parameterization of the $\tau_\Pi$ to be well-defined when $c_s^2 \geq 1/3$. The choices of these viscous relaxation times and other second-order transport coefficients are constrained by the causality conditions for relativistic viscous hydrodynamics~\cite{Bemfica:2020xym, Chiu:2021muk}.

In this work, we assume the following parameterization for the viscous relaxation times,
\begin{align}
    \tau_\pi &= C_\eta \frac{\eta}{(e + P)} = \frac{\eta}{s} \frac{C_\eta}{T}, \\
    \tau_\Pi &= C_\zeta \frac{\zeta}{(e + P)} = \frac{\zeta}{s} \frac{C_\zeta}{T},
\end{align}
where $C_\eta$ and $C_\zeta$ are constants, the second equality is only valid at zero chemical potential. Based on this parameterization, the linearized causality condition for second-order DNMR hydrodynamics can be written as~\cite{Hiscock:1983zz}
\begin{equation}
    0 \le c_s^2 + \frac{4}{3 C_\eta} + \frac{1}{C_\zeta} \le 1.
    \label{eq:causality}
\end{equation}
In this work, we set $C_\eta = 5$ and $C_\zeta = 5$ so that the constraints on the speed of sound squared in Eq.~\eqref{eq:causality} translates to
\begin{align}
    0 \le c_s^2 \le 1 - \frac{4}{3 C_\eta} - \frac{1}{C_\zeta} = \frac{8}{15}.
\end{align}
Therefore, we can ensure the linearized causality condition for any EOS with $c_s^2$ smaller than $1/2$.

According to Ref.~\cite{Denicol:2014vaa}, another second-order transport coefficient $\lambda_{\Pi \pi}$, which couples the bulk viscous pressure with the velocity shear tensor in the DNMR theory, has explicit dependence on the factor $(1/3 - c_s^2)$ as
\begin{align}
    \lambda_{\Pi \pi} = \frac{8}{5} \left(\frac{1}{3} - c_s^2\right) \tau_\Pi.
\end{align}
In this work, we set it to be the maximum value for any values of $c_s^2$,
\begin{align}
    \lambda_{\Pi \pi} = \frac{8}{15} \tau_\Pi.
\end{align}

\subsection{EOS effects on final-state hadronic observables}
\label{subsec:observables}
We will now examine a few final-state observables and the effect of different EOS on them. In this work, we keep all the model parameters fixed in the simulations and only change the EOS to show its effects on final-state observables.
These results can provide useful insights into which observables are highly sensitive to EOS and what temperature range of the EOS the final-state observables can probe.
Using the methods developed in Ref.~\cite{Roch:2024xhh}, we will perform a Bayesian inference study to constrain EOS together with all the other model parameters in future work because such a study requires large-scale numerical simulations.

Before we dive into the results, it is helpful for the understanding of the different behavior of the observables to note that the fireball spends most of its lifetime in the temperature region around $150-250$~MeV~\cite{Paquet:2015lta}, where we observe a clear ordering in the speed of sound squared: $c_s^2|_{\text{EOS 2}}<c_s^2|_{\text{HotQCD}}<c_s^2|_{\text{EOS 1}}$. The flow development in the hydrodynamic phase is primarily sensitive to the values of $c_s^2$ in this temperature region. 

\begin{figure*}[t!]
    \centering
    \includegraphics[width=0.49\textwidth]{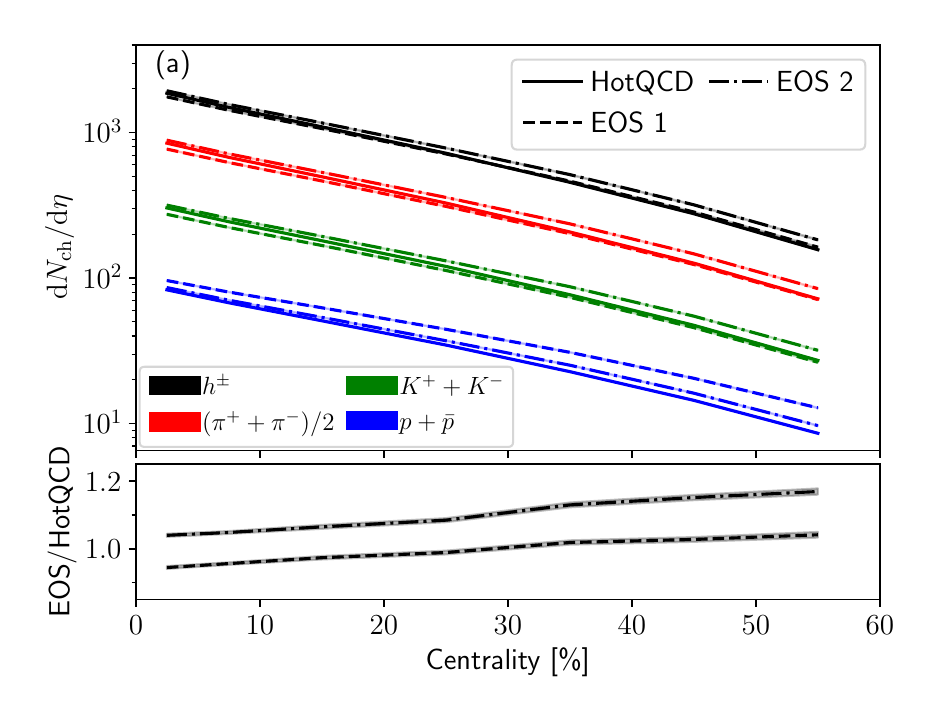}
    \includegraphics[width=0.49\textwidth]{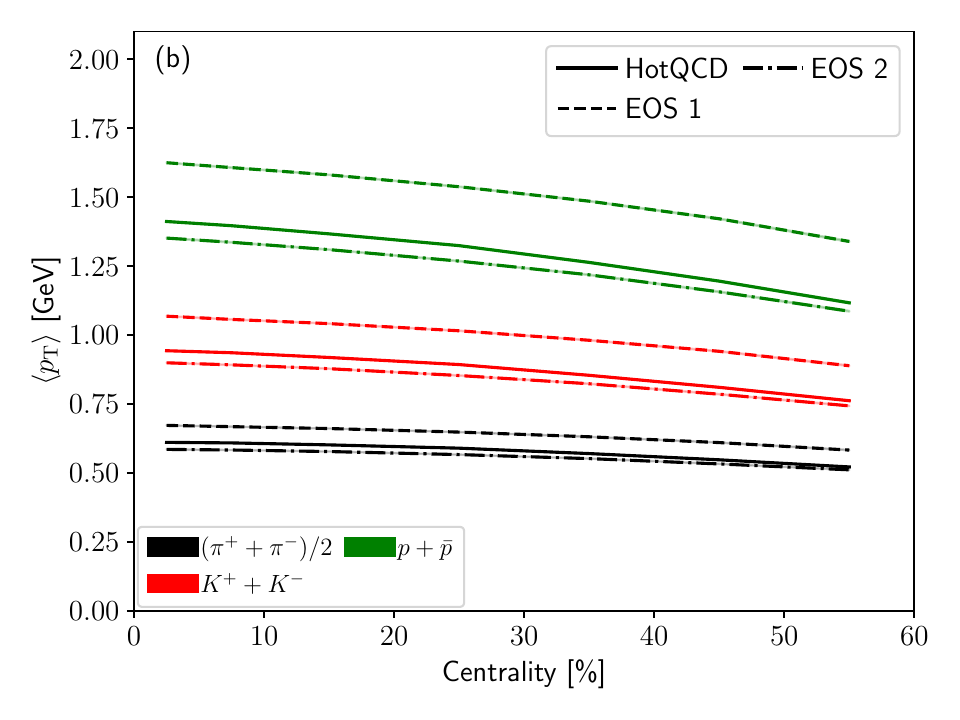}
    \includegraphics[width=0.49\textwidth]{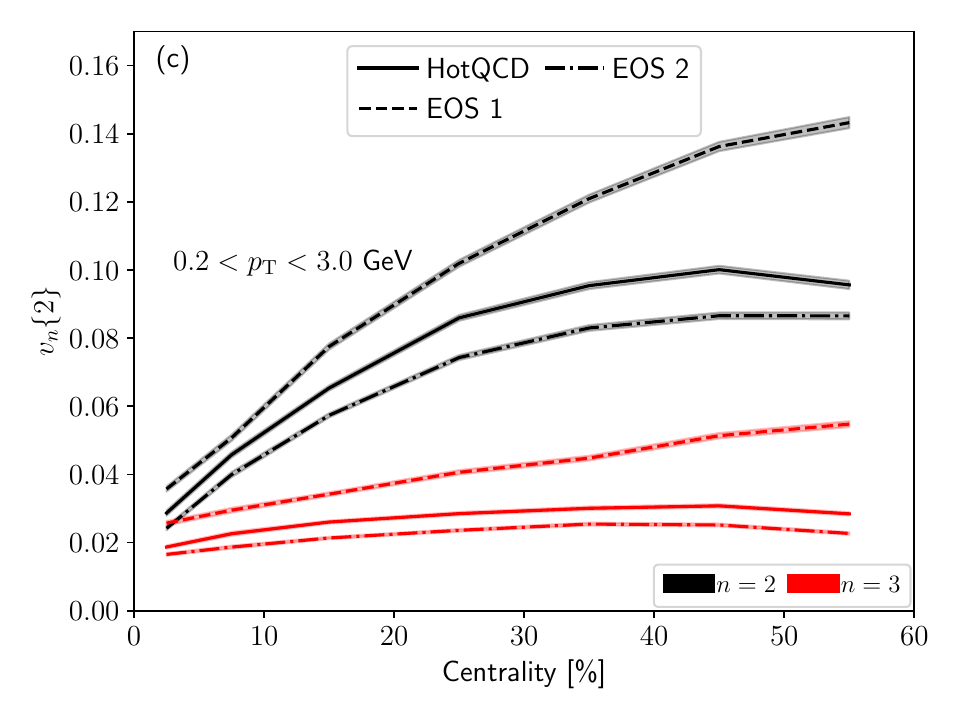}
    \includegraphics[width=0.49\textwidth]{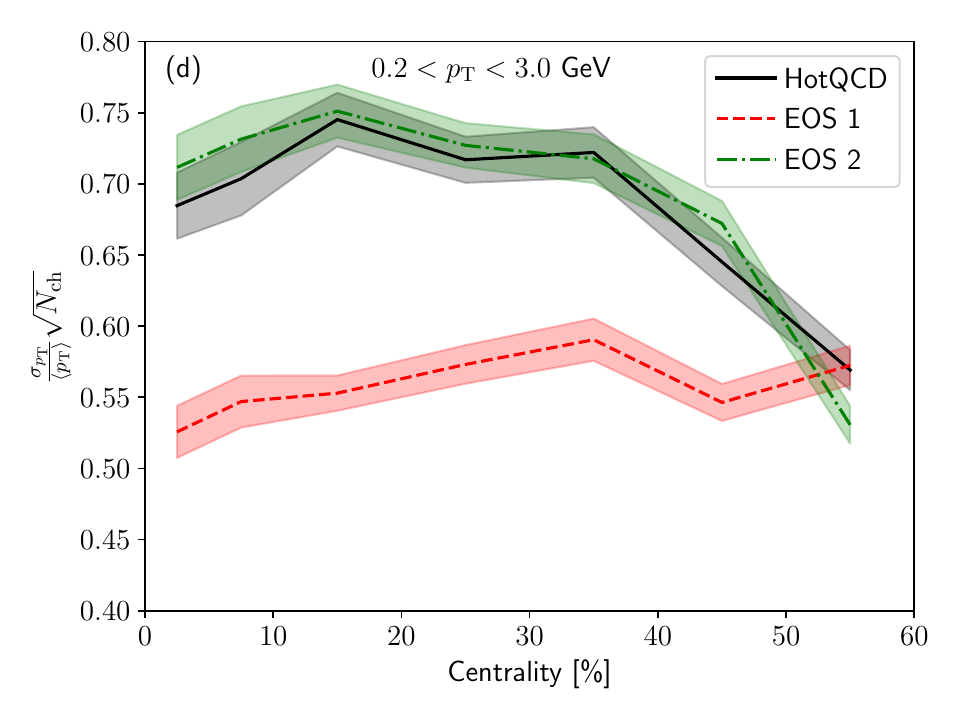}
    \caption{The centrality dependence of charged hadron and identified particle multiplicities (a), identified particle averaged transverse momenta $\langle p_{\rm T}\rangle$ (b), charged hadron anisotropic flow coefficients $v_n\{2\}\, (n=2, 3)$ (c), and the scaled standard deviation of charged hadron transverse momentum fluctuation (d) in Pb+Pb collisions at $\sqrt{s_\mathrm{NN}} = 5.02$~TeV using three different equations of state.}
    \label{fig:obs1}
\end{figure*}

In Fig.~\ref{fig:obs1}, we show the EOS effects on a few $p_{\rm T}$-integrated observables as functions of collision centrality.
The yields of charged hadron and identified particles in Fig.~\ref{fig:obs1}a show some sensitivity to the EOS.
Because the three EOS are constrained by the HRG and lattice QCD results at low and high temperatures, respectively, the initial total entropies of the collision systems from the three EOS only differ by $\sim15$\% from each other.
%This is because most of the system's entropy is produced in the IP-Glasma stage. The difference among the results from different EOS mainly comes from the different amounts of viscous entropy production during the hydrodynamic evolution.
We observe that EOS 1 gives a larger $p$ to $\pi$ ratio than those from the other two EOS sets, indicating significant bulk viscous corrections at the particlization because it has a large $c_s^2$ near the cross-over region, which leads to fast expansion.

During the hydrodynamic evolution, the speed of sound in the EOS controls the size of local acceleration of the fluid velocity~\cite{Heinz:2004qz},
\begin{align}
    {\rm D} u^\mu = \frac{\nabla^\mu P}{e + P} \approx \frac{c_s^2}{1 + c_s^2} \frac{\nabla^\mu e}{e}.
\end{align}
The larger the $c_s^2$, the stronger the acceleration from the same energy density gradients.

Figure~\ref{fig:obs1}b shows the identified particle average transverse momenta $\langle p_{\rm T}\rangle$ as functions of centrality from the three equations of state. Because final-state hadrons' mean $p_{\rm T}$ imprints the underlying hydrodynamic radial flow, hydrodynamic evolution with EOS 1 gives the largest mean $p_{\rm T}$ for identified hadrons among the three. The ordering among the three EOS follows closely with their $c_s^2$ values in the temperature region between 150 and 250 MeV.
The fast acceleration with the large speed of sound also results in larger anisotropic flow coefficients for charged hadrons, as shown in Fig.~\ref{fig:obs1}c. Again, we find the ordering of the results among the three equations of state is the same as their speed of sound near the cross-over region.

Since the speed of sound in equations of state is tightly correlated with the hydrodynamic flow event-by-event, we further study its effects on the fluctuation of hydrodynamic flow, which are imprinted in the charged hadron transverse momentum fluctuation. We compute the variance of the charged hadron $p_{\rm T}$ fluctuation with
\begin{align}
    \sigma_{p_{\rm T}}^2 = \left\langle\frac{\sum_{ij} \delta p_{{\rm T}, i} \delta p_{{\rm T}, j}} {N_i N_j} \right\rangle,
\end{align}
where the index $i$ sums over all charged hadrons in one subevent with $\eta \in [-0.8, -0.4]$ and $p_{\rm T} \in [0.2, 3]$ GeV and the index $j$ sums over all charged hadrons with $\eta \in [0.4, 0.8]$ and $p_{\rm T} \in [0.2, 3]$ GeV in the same event. The $p_{\rm T}$ fluctuation of each individual charged hadron is defined as $\delta p_{{\rm T}, i/j} \equiv p_{{\rm T}, i/j} - \langle p_{\rm T} \rangle$ with charged hadron averaged transverse momentum $\langle p_{\rm T} \rangle$ computed in the same kinematic cuts in the subevent. The denominator $N_i N_j$ counts the number of particle pairs in the event. The $\langle \cdots \rangle$ averages over collision events.

Figure~\ref{fig:obs1}d shows the normalized standard deviation of charged hadron $p_{\rm T}$ scaled by $\sqrt{N_\mathrm{ch}}$ as a function of centrality. Here, $N_{\rm ch}$ is the charged hadron multiplicity in $p_{\rm T} \in [0.2, 3]$ GeV and $\vert \eta \vert \in [0.4, 0.8]$. The factor $\sqrt{N_\mathrm{ch}}$ scales out the trivial centrality dependence of $p_{\rm T}$ fluctuations on finite numbers of charged hadrons~\cite{ALICE:2014gvd, Schenke:2020uqq}.
We observe that EOS 1 with a large speed of sound near the cross-over region results in small transverse momentum fluctuations. Furthermore, the centrality dependence of this scaled observable is also sensitive to the speed of sound in the EOS. The EOS with the low speed of sound in the temperature region $T \in [150, 250]$ MeV gives a decreasing trend of this scaled observable, while the results from EOS 1 are flat as a function of centrality.

\begin{figure}[th!]
    \centering
    \includegraphics[width=0.49\textwidth]{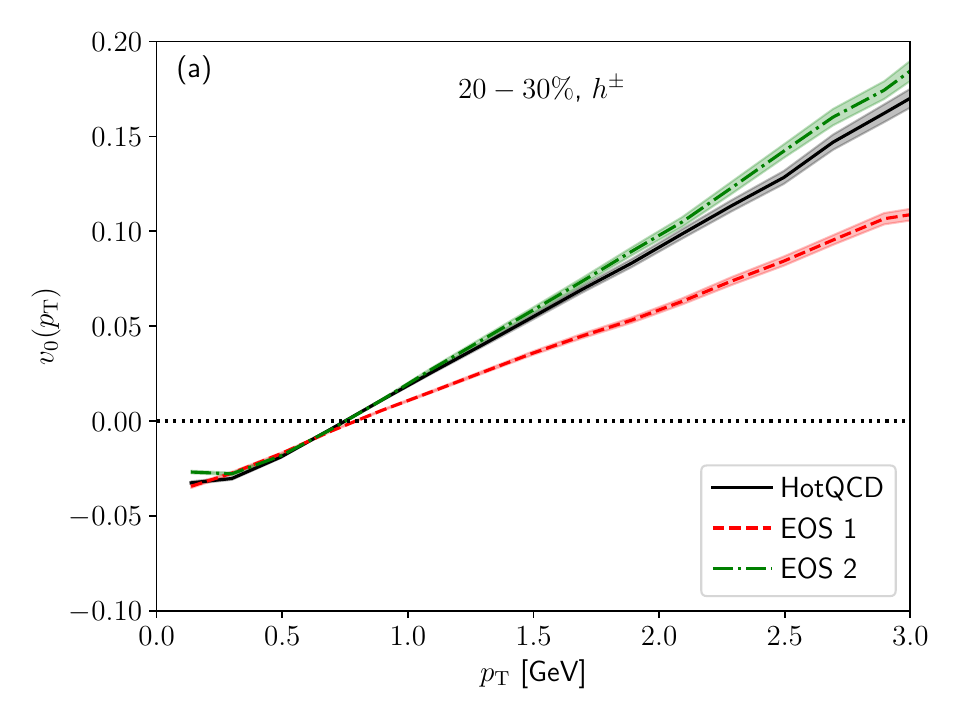}
    \includegraphics[width=\linewidth]{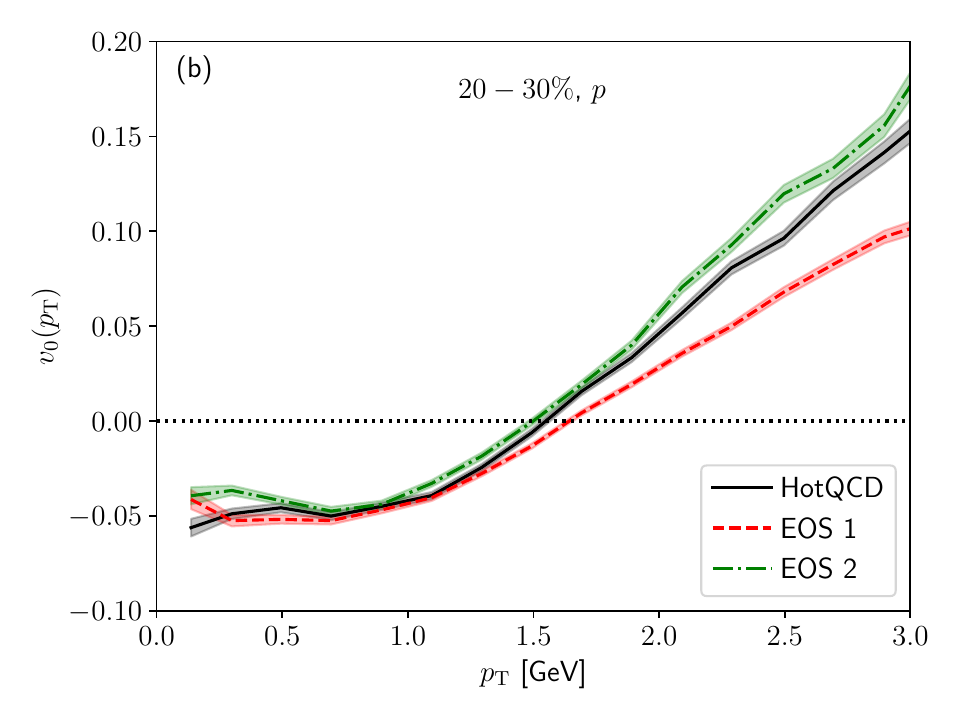}
    \caption{The $p_{\rm T}$-differential $v_0(p_{\rm T})$ for charged hadrons (a) and protons (b) at $20-30\%$ Pb+Pb collisions at $\sqrt{s_\mathrm{NN}} = 5.02$~TeV from three different equations of state.}
    \label{fig:obs2}
\end{figure}

The charged hadron $p_{\rm T}$ fluctuations can be further studied with the $v_0(p_{\rm T})$ observables proposed in Ref.~\cite{Schenke:2020uqq, Parida:2024ckk}. In a simple model~\cite{Gardim:2019iah, Schenke:2020uqq}, this observable can be interpreted as follows,
\begin{align}
    v_0(p_{\rm T}) \approx 2 \frac{\sigma_{p_{\rm T}}}{\langle p_{\rm T} \rangle} \left(\frac{p_{\rm T}}{\langle p_{\rm T} \rangle} - 1 \right).
    \label{eq:v0pT}
\end{align}
According to Eq.~\eqref{eq:v0pT}, $v_0(p_{\rm T})$ crosses zero when $p_{\rm T} = \langle p_{\rm T} \rangle$ and the slope of $v_0(p_{\rm T})$ as a function of $p_{\rm T}$ is related to the normalized standard deviation of the $p_{\rm T}$ fluctuations. Figure~\ref{fig:obs2} shows our simulation results of $v_0(p_{\rm T})$ for charged hadrons in panel (a) and for protons in panel (b). We observe that the numerical results agree very well with the interpretations based on Eq.~\eqref{eq:v0pT}. Proton's $v_0(p_{\rm T})$ cross zero at $p_{\rm T} \sim 1.5$ GeV, inline with their mean $p_{\rm T}$ values shown in Fig.~\ref{fig:obs1}b. Comparing $v_0(p_{\rm T})$ from the three equations of state, we find that the ordering of the $p_{\rm T}$ values when $v_0(p_{\rm T}) = 0$ follows the mean $p_{\rm T}$ ordering in Fig.~\ref{fig:obs1}b.
Simulations with EOS 1 have a smaller slope of $v_0(p_{\rm T})$ compared to those results from the other two EOS. It is consistent with the ordering of the $p_{\rm T}$ fluctuations shown in Fig.~\ref{fig:obs1}d.

\begin{figure*}[t!]
    \centering
    \includegraphics[width=0.8\textwidth]{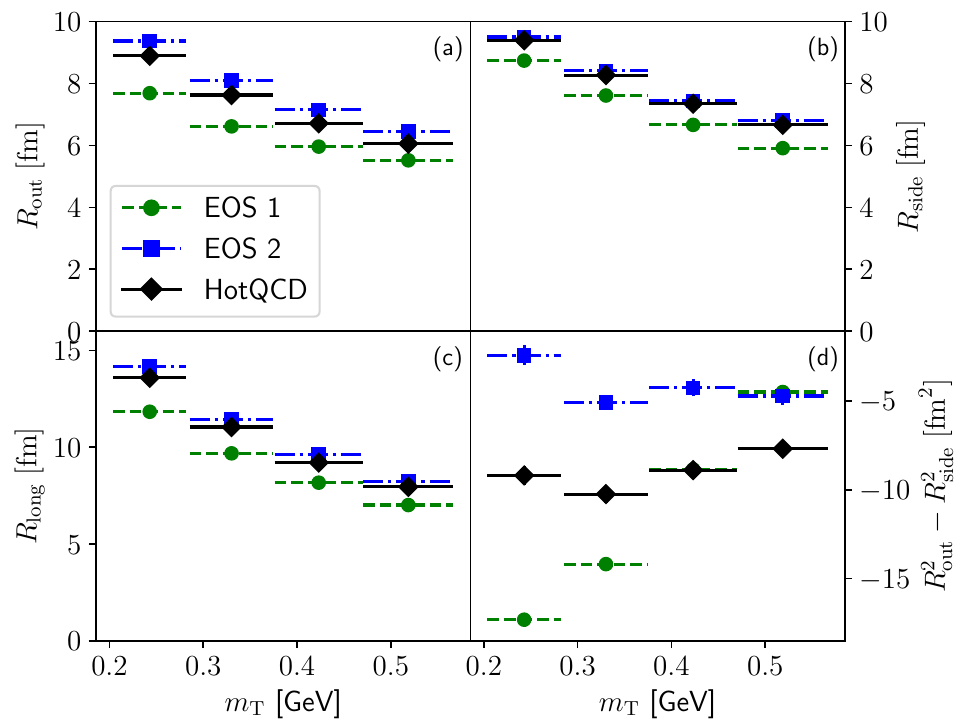}
    \caption{HBT radii $R_{\rm out}$, $R_{\rm side}$, $R_{\rm long}$ and $R_{\rm out}^2-R_{\rm side}^2$ of identical pion pairs as a function of $m_{\rm T}$ for the different equations of state in mid-rapidity $0-5\%$ Pb+Pb collisions at $\sqrt{s_\mathrm{NN}} = 5.02$~TeV.}
    \label{fig:HBT}
\end{figure*}

The two-particle Hanbury-Brown-Twiss (HBT) correlation is a sensitive probe for the speed of sound in the system~\cite{Heinz:2004qz, Lisa:2005dd, Pratt:2008qv, Pratt:2015zsa}. 
For pairs of identical pions in the final state with momenta $p^\mu_i$ and $p^\mu_j$, we define their pair momentum $K^\mu = (p^\mu_i + p^\mu_j)/2$ and momentum difference $q^\mu_{ij} = p^\mu_i - p^\mu_j$. Then, their HBT correlation can be computed as,
\begin{align}
    C(K_\mathrm{T}, \vec{q}) = 1 + \frac{\frac{1}{\langle N_{\rm pair} \rangle} \langle \sum_{ij} \cos(q_{ij} \cdot x_{ij}) \rangle}{\frac{1}{\langle N_{\rm mix pair} \rangle} \langle N_{\rm mix pair} (q) \rangle},
    \label{eq:HBTcorrelation}
\end{align}
where the indices $ij$ runs over all particle pairs $(i, j)$ within the transverse pair momentum $K_{\rm T} = \vert \vec{K}_{\rm T} \vert$ bin and individual rapidity $y_{i/j} \in [-0.5, 0.5]$.
The four-vector $x^\mu_{ij} = x^\mu_i - x^\mu_j$ with $x^\mu_i$ the space-time position of the last interaction point of particle $i$. The HBT correlation $C(K_\mathrm{T}, \vec{q})$ in Eq.~\eqref{eq:HBTcorrelation} is usually expressed in the out-side-long coordinate system for $\vec{q} = (q_\mathrm{out}, q_\mathrm{side}, q_\mathrm{long})$. We perform a fit to the HBT correlation function using a 3D Gaussian~\cite{STAR:2014shf},
\begin{align}
    C(K_\mathrm{T}, \vec{q}) &= \lambda \exp \left(- q^2_\mathrm{out} R_\mathrm{out}^2 - q_\mathrm{side}^2 R_\mathrm{side}^2 - q_\mathrm{long}^2 R_\mathrm{long}^2 \right. \nonumber \\
    & \qquad \qquad - 2 q_\mathrm{out} q_\mathrm{side} R_{\rm os}^2 - 2 q_\mathrm{out} q_\mathrm{long} R^2_{\rm ol} \nonumber \\
    & \qquad \qquad - 2 q_\mathrm{side} q_\mathrm{long} R^2_{\rm sl}),
\end{align}
from where we extract the HBT radii $R_{\rm out}, R_{\rm side}, R_{\rm long}$ together with the cross terms $R_{\rm os}, R_{\rm ol}, R_{\rm sl}$.

Figure~\ref{fig:HBT} shows the HBT radii $R_{\rm out}$, $R_{\rm side}$, $R_{\rm long}$, and $R_{\rm out}^2-R_{\rm side}^2$ as functions of pion's transverse mass $m_{\rm T} \equiv \sqrt{m_\mathrm{pion}^2 + K_\mathrm{T}^2}$ for $0-5\%$ Pb+Pb collisions at $\sqrt{s_\mathrm{NN}} = 5.02$~TeV.
We find that the region of homogeneity in all three directions, measured by $R_{\rm out}$, $R_{\rm side}$, and $R_{\rm long}$, shows the same trends with decreasing values for equations of state with a high speed of sound near the cross-over region. Because the larger $c_s^2$ leads to faster hydrodynamic expansion in the transverse plane, the fireball has a shorter lifetime to grow its size when most of the pions are emitted, resulting in a smaller $R_\mathrm{side}$ in panel (b)~\cite{Pratt:2008qv}. For $R_\mathrm{out}$ and $R_\mathrm{long}$, they are sensitive to the time duration of the pion emission in addition to the spatial size of the homogeneous emission region in the fireball~\cite{Wiedemann:1999qn}. The shorter fireball lifetime with EOS 1 results in smaller HBT radii in these two directions.

In Fig.~\ref{fig:HBT}d, we show the difference between $R^2_\mathrm{out}$ and $R^2_\mathrm{side}$, which is sensitive to the averaged transverse flow velocity and lifetime of the fireball~\cite{Wiedemann:1999qn}. With all three equations of state, $R_\mathrm{out} < R_\mathrm{side}$ in all $m_{\rm T}$ bins. We can understand this trend based on a simplified Gaussian emission source~\cite{Heinz:2004qz},
\begin{align}
    R^2_\mathrm{side} &= \langle x_{\rm s}^2 \rangle, \\
    R^2_\mathrm{out} & = \langle x_{\rm o}^2 \rangle - 2 \beta_\perp \langle x_{\rm o} t \rangle + \beta_\perp^2 \langle t^2 \rangle,
\end{align}
where $\langle x_{\rm s/o}^2 \rangle$ is the averaged fireball size squared in the side and out directions, respectively. 
For central Pb+Pb collisions, $\langle x_{\rm o}^2 \rangle - \langle x_{\rm s}^2 \rangle \rightarrow 0$ when $K_{\rm T} \rightarrow 0$.
Our results in Fig.~\ref{fig:HBT}d indicate that the space-time correlation $\langle x_{\rm o} t \rangle$ on the freeze-out surface is significantly positive in $0-5\%$ Pb+Pb collisions at $\sqrt{s_\mathrm{NN}} = 5.02$~TeV. With a larger average transverse velocity $\beta_\perp$ in simulations with EOS 1, the values of $R^2_\mathrm{out}-R^2_\mathrm{side}$ are more negative at small $m_{\rm T}$. We find $R^2_\mathrm{out}-R^2_\mathrm{side}$ at small $m_{\rm T}$ shows strong sensitivity to the speed of sound in the equation of state. Furthermore, the  $m_{\rm T}$ dependence of $R^2_\mathrm{out}-R^2_\mathrm{side}$ is qualitatively different among the three equations of state.

Looking back at the EOS effects on final state observables in Figs.~\ref{fig:obs1}, \ref{fig:obs2}, and \ref{fig:HBT}, we find that all these hadronic observables are sensitive to the speed of sound near the crossover temperature region, $T \in [0.15, 0.25]$~GeV. The EOS 2 has the highest speed of sound when $T \in [0.3, 0.4]$~GeV; however because only a small fraction of fluid cells probes this temperature region such that hadronic observables are not sensitive to them.

\subsection{EOS effects on electromagnetic radiation}

Electromagnetic probes, such as thermal photons and dileptons, are sensitive to the local temperature and flow field at their emission points. They carry valuable information about early-time dynamics of heavy-ion collisions, complementary to final-state hadronic observables~\cite{Shen:2016odt, Gale:2018vuh, Gale:2021emg, Shen:2023aeg, Churchill:2023zkk, Wu:2024pba}.

\begin{figure}[t!]
    \centering
    \includegraphics[width=\linewidth]{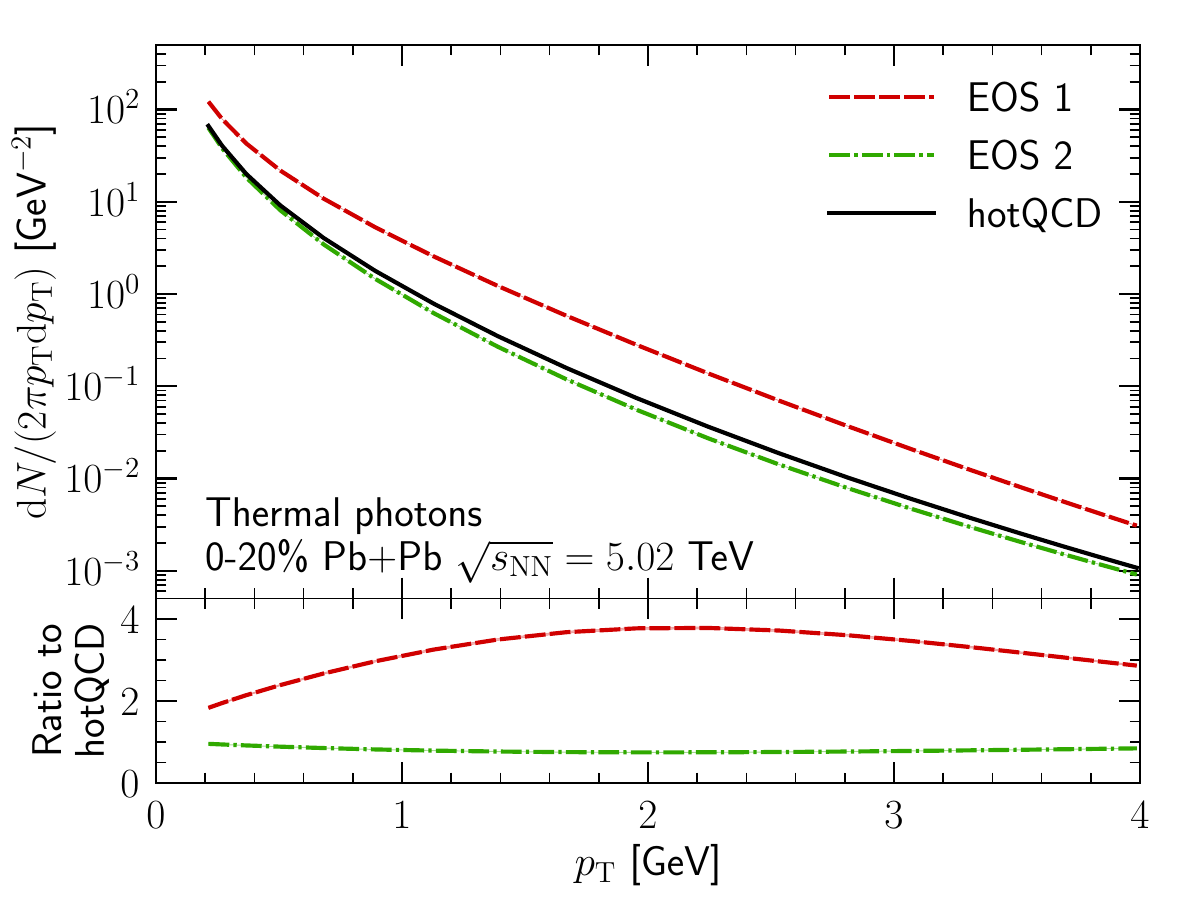}
    \caption{Thermal photon (QGP + hot hadron gas) $p_{\rm T}$-spectra from three equations of state in $0-20\%$ Pb+Pb collisions at $\sqrt{s_\mathrm{NN}} = 5.02$~TeV.}
    \label{fig:photonSP}
\end{figure}

Figure~\ref{fig:photonSP} shows the thermal photon spectra emitted from QGP and hot hadron gas phases in $0-20\%$ Pb+Pb collisions with the three different equations of state. They are computed by convoluting the thermal photon emission rates with the fireballs' dynamical evolution profiles event by event. We used the leading-order QGP photon emission rates~\cite{Arnold:2001ms} for fluid cells with temperatures above 180 MeV and hadronic gas photon emission rates~\cite{Turbide:2003si, Heffernan:2014mla} for those fluid cells with temperatures between 120-180 MeV. The leading order out-of-equilibrium corrections to thermal photon emission rates are included for the available photon emission channels~\cite{Shen:2014nfa, Paquet:2015lta, Hauksson:2016nnm}.

We observe that the simulations with EOS 1 result in a factor of 3 more thermal photons than those from simulations with the other two equations of state. The EOS 1 has the smallest $P/T^4$ values among the three equations of state, as shown in Fig.~\ref{fig:EOS}, which leads to the highest local temperature for fluid cells at a given energy density. Then, the higher temperature in EOS 1 results in more thermal photon radiation. Unlike the hadronic observables, which are sensitive to the time-integrated EOS effects on hydrodynamic flow, the thermal photon spectra show direct sensitivity to the temperature profile of the collisions given by the equation of state. Figure~\ref{fig:photonSP} demonstrates that the yields of thermal photon radiation are a direct probe to the thermodynamic properties of the hot QCD matter.

\begin{figure}[h!]
    \centering
    \includegraphics[width=\linewidth]{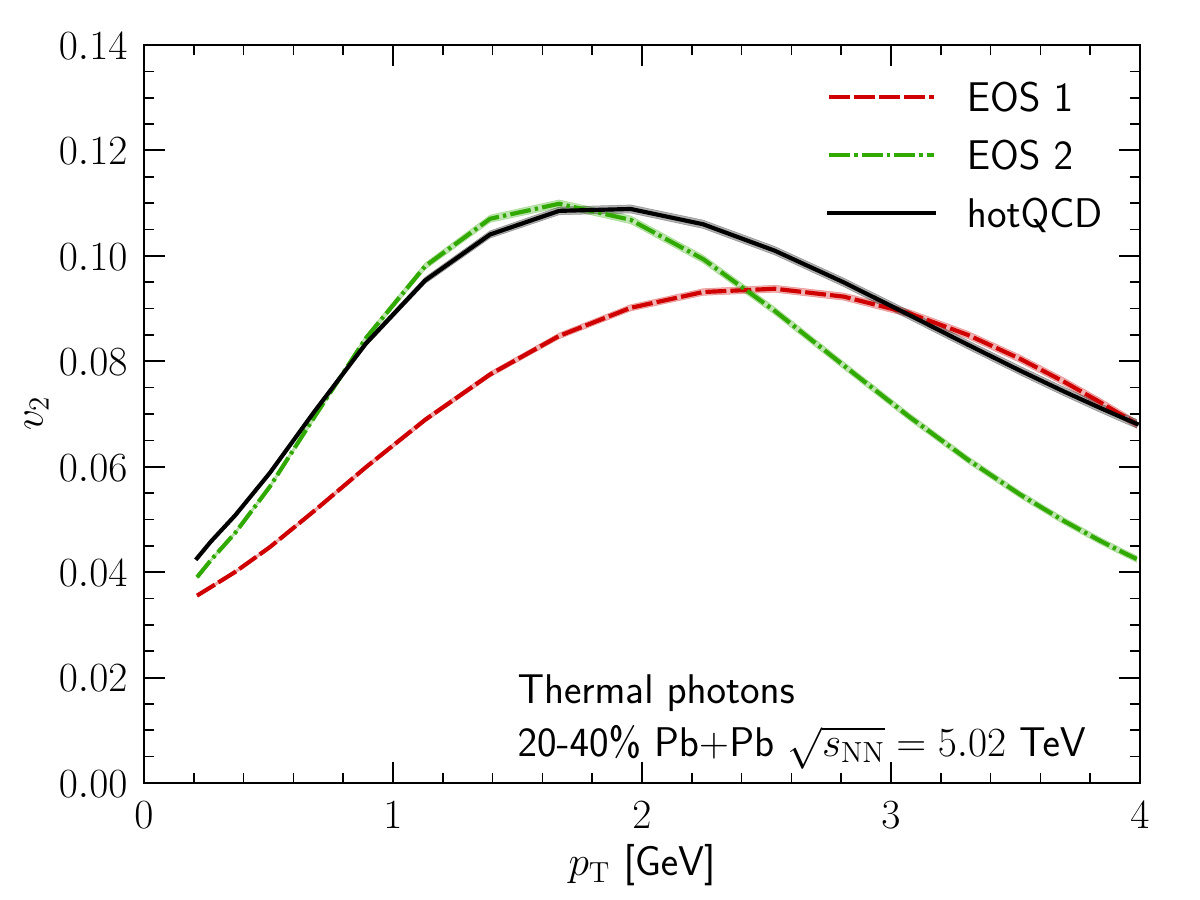}
    \caption{Thermal photon elliptic flow coefficient for $20-40\%$ Pb+Pb collisions at $\sqrt{s_\mathrm{NN}} = 5.02$~TeV with three different equations of state.}
    \label{fig:photonV2}
\end{figure}

The fireball's flow profile can be probed by the elliptic flow coefficients of thermal photons, as shown in Fig.~\ref{fig:photonV2}. Different from the hadrons' elliptic flow, thermal photons with high transverse momenta are mostly emitted during the early phase of the evolution at which the hydrodynamic flow has not fully developed. It results in the rise and fall structure in the $p_{\rm T}$-dependence of thermal photon elliptic flow coefficients. We find that the simulations with EOS 1 have the lowest $v_2$ for $p_{\rm T} < 2$~GeV among the three equations of state. The trend in thermal photon $v_2(p_{\rm T})$ is the exact opposite of the ordering in charged hadrons $v_2$ shown in Fig.~\ref{fig:obs1}c. Because EOS 1 has the highest temperatures at a given energy density compared to those from the other two equations of state, the thermal photon radiation from EOS 1 is dominated by the high-temperature phase where the hydrodynamic anisotropic flow has not fully developed. There are more photon emissions from the hot hadron gas phase with EOS 2 and the hot QCD EOS, which increases the thermal photon $v_2$ in the low $p_{\rm T}$ region.

\section{Conclusion}
\label{sec:conclusion_outlook}
In this paper, we employ Gaussian Process regression as a generative model to produce random smooth cross-over type QCD equations of state. This model allows us to impose desired physics constraints from thermodynamics and causality and provides enough variations in the allowed phase space. We demonstrate the results at zero baryon density, where the thermal pressure is only a function of temperature $P(T)$. Our formulation can easily be extended to higher dimensions, including dependencies on conserved charges like baryon, strangeness, and electric charges. 

We deploy this EOS generator to the \texttt{iEBE-MUSIC} framework and perform event-by-event simulations for Pb+Pb collisions at $\sqrt{s_\mathrm{NN}} = 5.02$~TeV to explore the effects of EOS on final-state observables. 
We find that most of the hadronic observables are sensitive to the values of the speed of sound near the cross-over region between the hadron gas and QGP phases, where $T \in [0.15, 0.25]$~GeV. A high speed of sound in the equation of state results in a large acceleration rate from the local energy density gradients and leads to the fast development of the radial and anisotropic flow during the hydrodynamic phase. 
Hadron's averaged transverse momenta, the variance of $p_{\rm T}$ fluctuations, and anisotropic flow coefficients show strong responses to the variations in EOS. The HBT radii of identical pion pairs are also sensitive observables for probing the speed of sound in the equation of state. 
Finally, we study the effects of the EOS on electromagnetic radiation. We find that the yields of thermal photons are strongly affected by the temperature profile encoded in the equation of state. It demonstrated the tomographic nature of thermal photons to probe the thermodynamic properties of hot QCD matter.
The elliptic flow of thermal photons is sensitive to the relative photon emission at different temperatures and times, providing complementary information about the hydrodynamic flow development to those inferred from hadrons' anisotropic flow coefficients.

The observed strong sensitivities in final-state observables to the variations in the equation of state pave a good starting point for a systematic Bayesian inference study on constraining the QCD EOS using experimental measurements from relativistic heavy-ion collisions. It would be interesting to compare the results from such a data-driven approach at zero baryon density with the ones from the first-principle calculation, which we will pursue as future work.
We also plan to extend the current generative model to finite baryon density and carry out dedicated Bayesian inference studies to constrain the behavior of QCD EOS at large baryon density with measurements in the RHIC Beam Energy Scan program.

The code developed for this work is open source and
can be downloaded from~\cite{gong_2024_14008389}.

\begin{acknowledgments}
This work is supported in part by the U.S. Department of Energy, Office of Science, Office of Nuclear Physics, under DOE Award No.~DE-SC0021969 and DE-SC0024232.
H.~R. was supported in part by the National Science Foundation (NSF) within the framework of the JETSCAPE collaboration (OAC-2004571).
C.S. acknowledges a DOE Office of Science Early Career Award. 
This research was done using computational resources provided by the Open Science Grid (OSG)~\cite{Ruth_Pordes_2007,5171374}, which is supported by the National Science Foundation award \#2030508 and \#1836650.
\end{acknowledgments}

\bibliography{bib, non-inspire}

\end{document}